\documentstyle[aps,prl,epsfig,twocolumn]{revtex}

\def\be{\begin{equation}}
\def\ee{\end{equation}}
\def\bea{\begin{eqnarray}}
\def\eea{\end{eqnarray}}
\def\bma{\begin{mathletters}}
\def\ema{\end{mathletters}}
\def\C{\hbox{$\mit I$\kern-.6em$\mit C$}}

\begin{document}
\draft

\title{Quantum Measured Information}

\author{ Yi-Xin Chen}

\address{Zhejiang Institute of Modern Physics and
Department of Physics, Zhejiang University, Hangzhou 310027, P.R. China} 

\date{\today}

\maketitle

\begin{abstract}
A framework for a quantum information theory is introduced that is based on the
measure of quantum information associated with probability distribution 
predicted by quantum measuring of state. The entanglement between states of 
measured
system and "pointer" states of measuring apparatus, which is generated by
dynamical process of quantum measurement, plays a dominant role in expressing
quantum characteristics of information theory. The quantum mutual information
of transmission and reception of quantum states along a noisy quantum channel
is given by the change of quantum measured information. In our approach, it is
not necessary to purify the transmitted state by means of the reference system.
It is also clarified that there exist relations between the approach given in 
this letter and those given by other authors.
\end{abstract}

\pacs{03.65.Bz,  89.70.+c}

\narrowtext


Quantum information theory is a new field with potential implication for
the conceptual foundations of quantum mechanics. It appears to be the base
for a proper understanding of the emerging fields of quantum computation
\cite {Shor,Bennet,DiVincenzo,Lloyd}, quantum communication, and quantum
cryptography\cite {Bennet,Bennet2}.
Recently, a correlated state in quantum systems, so-called quantum
entangled or quantum entanglement, is utilized to study quantum information
\cite {Schumacher,Barnum}, in particular, quantum teleportation.
A theory of quantum information has emerged which shows striking parallels with,
but also fascinating differences from, classical information theory entirely
based on the von Neumann entropy of quantum state\cite {Adami}. Although
some useful fundamental results about quantum information theory, e.g., quantum
noiseless coding theorem\cite{Schumacher1} and the capacity of
quantum noisy channels\cite{Barnum,Adami}, have been obtained recently,
quantum information is still mystery in many respects. To our knowledge, there
exist several approaches for quantum information theory 
\cite {Schumacher,Adami,Ohya}.
However, the relation among these approaches is obscure. We argue in this 
letter that it is necessary to introduce quantum measured information 
which is regarded as the measure of information for quantum input and output. 
This leads us to propose a scheme for quantum information theory. Our
approach can give rise to a unified description of classical correlation and
quantum entanglement. Furthermore, we shall clarify the relations between our
approach and those given by other authors for quantum information theory.

In classical information theory, there is a set of mutually exclusive classical
states. In quantum mechanics, a quantum state is represented by a vector in a
Hilbert space, or a density operator on that space. Classically, the input
system may retain its original state, while the no-cloning theorem
\cite{Wootters}
implies that in the quantum case the input system cannot in general remain in
its initial state. However, in many quantum applications, one is interested 
not only in transmitting a discrete set of states, but also in arbitrary
superpositions of those states. That is, one wants to transmit entire subspace
of states. It is well known that an arbitrary state can be represented as a
mixture of pure states, i.e., by imposing classical randomness on pure states.
In this sense pure states are "noiseless", i.e., they contain no classical
source of randomness. In our point of view, this does not imply that the pure 
states
contain no quantum source of randomness. Such randomness comes from probability
distribution predicted by quantum measuring about quantum states.

We now consider the question of measurement in quantum mechanics. From our 
point of view there is no fundamental distinction between measuring apparatus 
and
other physical systems. Therefore, a measurement is simply a special case of
interaction between physical systems, which has the property of correlating a
quantity in one system with a quantity in the other. Nearly every interaction
between systems produces some correlation however. Suppose that at some instant
a pair of systems are independent, so that the composite system state function
is a product of subsystem states. Then this condition obviously holds only
instantaneously, since the systems are interacting, the independence will be 
immediately destroyed and the systems will become correlated. There is still 
one
more requirement that we must impose on an interaction before we shall call it
a measurement. If the interaction is to produce a measurement of mechanical
quantity $A$ of subsystem $S_{1}$ by the quantity $P$ of another one $S_{2}$,
we require that such interaction shall never decrease the information in the
reduced distribution about $A$. Furthermore, we also expect that a knowledge of
$P$ shall give us more information about $A$ than we had before the measurement
took place, since otherwise the measurement would be useless. The restriction
that the interaction shall not decrease the information of the reduced system
$S_{1}$ has the interacting consequence that the eigenstates of $A$ will not be
disturbed, since otherwise the information of $A$ would be decreased. The time
evolution of the composed system made of  the measured system and the
apparatus system should exhibit the following properties. The initial states of
systems
$S_{1}$ and $S_{2}$ are the form of superposition $\sum_{i} a_{i}|\phi_{i}>$
and $|P(0)>$, respectively. Then, after a specified time of interaction the
total state $\sum_{i} a_{i}|\phi_{i}>|P(0)>$ will be transformed into a
a form of the superposition states $\sum_{i} a_{i}|\phi_{i}>| P_{i}>$,
i.e., the initial independent state is evolved into entanglement state.

Information concepts have been used in the context of quantum measurements long
ago\cite{Neumann}, and various quantities, all labeled entropies, have been
introduced to characterize uncertainties about events or about states of a
system\cite {Wehrl}. For a completed orthonormal set $\{ |\phi_{i}>\}$ and a
pure state $|\psi>=\sum_{i} a_{i}|\phi_{i}>$, we have a square-amplitude distribution $|a_{i}|^{2}$ called the distribution of $|\psi>$ over $\{|\phi_{i}>\}$.
In the probabilistic interpretation this distribution represents the 
probability distribution over the results of a measurement with eigenstates $\{|\phi_{i}>\}$ performed upon the measured system in the state $|\psi>$. An 
entropy of information depends not only on the actual set of probabilities 
for the considered
events, but also  on the measure associated with the {\it a priori}
probabilities predicted. Such entropy of information is given by Shannon entropy\cite{Shannon} $S=-\sum_{i} |a_{i}|^{2}\log |a_{i}|^{2}$. Considering the
dynamical process of quantum measurement as mentioned above, we can arrange
the distribution corresponding to the reduced density matrix of measured system
into the {\it a priori} probabilistic distribution predicted by quantum 
measurement.
While for the composite system made of the measured system and the apparatus 
system, its state is evolved into such a state $|\psi>=\sum_{i}a_{i}|\phi_{i}>|P_{i}>$. The reduced density matrix can be obtained by tracing
the state out the degrees of freedom of the apparatus system, i.e., $\rho_{M}=
Tr_{P}(|\psi><\psi|)=\sum_{i}|a_{i}|^{2}|\phi_{i}><\phi_{i}|$. The quantum measured information carried by the quantum state $|\psi>$ can be read as
\be
S_{M}(\rho)=S(\rho_{M})=-Tr\rho_{M}\log\rho_{M}.
\ee

In order to investigate the quantum mutual information, we shall consider the
mathematical description of model about a noisy quantum channel following
Schumacher\cite {Schumacher}. Suppose a quantum system $S$ is subjected to a
dynamical evolution, which may represent the transmission of $S$ along a noisy
quantum channel. In general, the evolution of $S$ will be represented by a
superoperator $\cal {E}$ which gives the mapping from the initial
states of the system $\rho$ to the final states after the evolution of the
system $\rho^{\prime}$, i.e., $\rho^{\prime}=\cal {E}(\rho)$. The
mapping represented by $\cal {E}$ is a linear, trace-nonincreasing and
completely positive map. The evolution of system will be unitary only if it is
isolated from other systems. The input quantum state, after interaction with an
environment, is lost, having become the output state. Any attempt at copying 
the quantum state before decoherence will result in a classical channel. Thus a
joint probability for input and output symbols does not exist for quantum
channels. However, this is not essential, as the quantity of interest is the
quantum measured information associated with the probabilistic distribution
predicted by quantum measurement, which is regarded as a measure of information
carried by a quantum state. Let us recall the definition of the mutual
information in classical information theory. The mutual information is a
measure of an amount of information that one random variable contains about 
another random variable, and is the reduction in the uncertainty of one 
random variable
due to the knowledge of the other. In the quantum case, the corresponding
quantity is the entropy of information given by that the change of quantum
measured distribution, which is resulted in by the evolution along a 
noisy quantum
channel. After the interaction with the environment, the quantum measured
distribution $\rho_{M}$ of quantum state becomes $\rho_{M}^{\cal {E}}={\cal E}(\rho_{M})=\sum_{i} |a_{i}|^{2}{\cal E}(|\phi_{i}>
<\phi_{i}|)$. In fact, The $\rho_{M}^{\cal {E}}$ can be equivalently
expressed as $\rho_{M}^{\cal {E}}=Tr_{P}{\cal E}\otimes
{\bf 1}_{P}(|\psi><\psi|)$. It should be noticed that the tracing process
implies the determination of probabilistic distribution predicted by quantum
measurement. In general, the state $\rho_{E}^{\cal E}={\cal E}\otimes
{\bf 1}_{P}(|\psi><\psi|)$ is a quantum mixing state, which carries the quantum
information given by the von Neumann entropy, i.e., $S(\rho_{E}^{\cal E})=-Tr\rho_{E}^{\cal E}\log\rho_{E}^{\cal E}$. In physics, it is known that the entropy 
change in the quantum states
represents the amount of the information obtained by the quantum measurement.
The amount of information of gain is equal to subtracting the information
$S(\rho_{E}^{\cal E})$ before quantum measurement from the quantum 
measured information
$S(\rho_{M}^{\cal {E}})$. This leads to the quantum mutual information
represented by
\be
I_{\cal {E}}=S(\rho_{M}^{\cal {E}})-S(\rho_{E}^{\cal E}),
\ee
based on the quantum measured information.

Up to now, we have discussed how the quantum mutual information is measured
when one transmits a pure quantum state 
along a noisy quantum channel by means of
concept of quantum measured information. It is well known that in general case
we should study the problem of transmission of some quantum mixed states 
because the states of a quantum system are fragile. Based on the 
following propositions
\cite {Wehrl}, we can easily generalize the above idea to the case of quantum
mixed states. The first proposition is that if $\rho$ is a pure state, there
exists a composite system made of two subsystems of which $\rho$ is the state
and the von Neumann entropies of the subsystems satisfy $S(\rho_{1})=S(\rho_{2})$, where $\rho_{1}$ and $\rho_{2}$ are the density matrixes of the subsystems.
Moreover, the positive spectra of $\rho_{1}$ and $\rho_{2}$ coincide. Secondly,
given $\rho_{1}$, one can always find a Hilbert space ${\cal {H}}_{2}$ and a 
pure
density matrix $\rho$ in the Hilbert space ${\cal {H}}_{1}\otimes {\cal {H}}_{2}$
such that $\rho_{1}=Tr_{2}\rho$. Now, let us suppose that a transmitted state
in the quantum mixed state $\rho_{1}=\sum_{m}p_{m}|S_{m}^{1}><S_{m}^{1}|$. We
can take an auxiliary Hilbert space ${\cal {H}}_{2}$ of which the dimensions 
are
the same as those of ${\cal {H}}_{1}$. Thus, a pure state of the composite 
system can
be constructed as $\rho=|\chi><\chi|$, here $|\chi>=\sum_{m}\sqrt {p_{m}}|S_{m}^{1}>|S_{m}^{2}>$. If one plans to measure the quantum mechanical quantity $A_{1}$ of the subsystem being in the state $\rho_{1}$, he should expand the state
$|\chi>$ with the quantity $A_{1}$ corresponding to the complete and orthogonal
eigenstates $\{ |\phi_{i}^{1}>\}$
\be
|\chi> =\sum_{m,i}\sqrt {p_{m}}c_{mi}|\phi_{i}^{1}>|\phi_{i}^{2}>
       =\sum_{i}\tilde {c}_{i}|\phi_{i}^{1}>|\phi_{i}^{2}>.
\ee

The dynamical evolution of quantum measurement is that of combining the 
measured system with the apparatus system of measuring, which leads to the state $|\chi>
|P(0)>$ into $|\tilde {\psi}>=\sum_{i}\tilde {c}_{i}|\phi_{i}^{1}>|\phi_{i}^{2}>
|P_{i}>$. So the quantum measured information of the quantum state $\rho_{1}$ 
is read as
\be
\begin{array}{l}
S(\tilde {\rho}_{M}) =-\sum_{i}|\tilde {c}_{i}|^{2}\log |\tilde {c}_{i}|^{2}
                    = -Tr \tilde {\rho}_{M}\log \tilde {\rho}_{M}
                    \cr ~~~~~~~~~
                     =-Tr [ Tr_{P}(|\tilde {\psi}><\tilde {\psi}|)\log Tr_{P}
                      (|\tilde {\psi}><\tilde {\psi}|) ],
\end{array}
\ee
where $\tilde {\rho}_{M}=Tr_{P,2}(|\tilde {\psi}><\tilde {\psi}|)=Tr_{\tilde
{P}}\sum_{i,j} \tilde {c}_{i}\tilde {c}_{j}^{*}|\phi_{i}^{1}>|\tilde {P}_{i}>
<\tilde {P}_{j}|<\phi_{j}^{1}|$. The set of states 
$\{ |\tilde {P}_{i}>= |\phi_{i}^{2}>|P_{i}>\}$
exhibits that the states of the apparatus system are completely entangled with
those of the auxiliary subsystem $S_{2}$. Hence, we can regard the states
$\tilde {P}_{i}$ as the "pointer" basis of quantum measurement. This implies
that the subsystem $S_{2}$ is completely auxiliary, and can be absent in our
formalism of quantum information theory. Consequently, it is not necessary in
our approach of quantum information theory based on the quantum measured
information to purify the initial transmitted state in the Schumacher's
approach\cite {Schumacher}. By means of the previous discussion
about the quantum mutual information of transmission of quantum pure state and
the corresponding reception, we can immediately write the expression of quantum mutual
information of transmission of quantum mixed states in the noisy quantum 
channel mentioned above. The result is
\begin{equation}
\tilde {I}_{\cal {E}}=S(\tilde {\rho}_{M}^{\cal {E}})
                         -S(\tilde {\rho}_{E}^{\cal {E}}).
\end{equation}
$\tilde {\rho}_{E}$ denotes the dynamically evolved state of the composite
system made of the measured system and the apparatus system, i.e., $\tilde {\rho}_{E}=|\tilde {\psi}><\tilde {\psi}|$. Through the noisy quantum channel, the
states $\tilde {\rho}_{E}$ and $\tilde {\rho}_{M}$ become $\tilde {\rho}_{E}^{
\cal {E}}={\cal {E}}\otimes {\bf 1}_{\tilde {P}}(\tilde {\rho}_{E})$ and $\tilde {\rho}_{M}^{\cal {E}}=Tr_{\tilde {P}}\tilde {\rho}_{E}^{\cal {E}}$, 
respectively. It should be emphasized again that the
$\tilde {\rho}_{M}^{\cal {E}}$ stands for the change of the
probabilistic distribution predicted by quantum measurement after the 
transmitted state going through the noisy quantum channel.

If a quantum channel is trivial, i.e., ${\cal {E}}=$ identity map, then
the quantum mutual information equals to the quantum measured information of
inputs. This is easily seen from the relation that $\tilde {I}_{\cal {E}}\mid_{\cal {E}=\bf {1}}=S(\tilde {\rho}_{M}^{\cal {E}})
\mid_{\cal {E}=\bf {1}}=S(\tilde {\rho}_{M})$. According to the
Araki-Lieb triangle inequality about the entropy of information\cite {Wehrl}, we
can lead to the quantum mutual information presented here satisfying $\tilde {I}_{\cal {E}}\leq S(\tilde {\rho}_{M})$ which is the first part of the
data processing inequality. Now, we shall consider a more complicated quantum
channel. Suppose the initial state of the measured system $S$ is $\rho_{S}$ and
further suppose $S$ undergoes two successive dynamical evolutions described by superoperators ${\cal {E}}_{1}$ and ${\cal {E}}_{2}$. This
scheme of the noisy quantum channel can be represented by the evolutions of
quantum state $\rho_{S}\rightarrow {\cal {E}}_{1}(\rho_{S})\rightarrow
{\cal {E}}_{2}\circ{\cal {E}}_{1}(\rho_{S})$. Following Schumacher and
Neilson\cite {Schumacher} and applying the strong subadditivity inequality, we
can prove the second part of the data processing inequality $\tilde {I}_{{\cal {
E}}_{1}}\geq \tilde {I}_{{\cal {E}}_{2}\circ{\cal {E}}_{
1}}$. 

On the other hand, by writing the superoperator $\cal {E}$ as a
unitary evolution $U_{SE}$ on an extended system $SE$ followed by a partial
trace over
an environment system $E$, we can investigate the reverse data processing
inequality in the quantum information theory, which reflects the fact that any
quantum channel used in a forward manner can be used in a backward manner. If 
we consider the case of transmission of quantum states in a backward manner 
and exchange
the input state with the output state, the order of time in the unitary
evolution $U_{SE}$ is reverse, which leads to the evolution operator to be
changed into $U_{SE}^{\dagger}$. So, in the transmission of backward manner, we
should substitute the evolution of state ${\cal {E}}^{\dagger}(\rho_{S})$ for 
the $\cal {E}(\rho_{S})$. Since the "pointer" state of apparatus
 system is completely entangled with the input states, the exchange between the
input states and the output states is equivalent to the change of 
$|\phi_{i}^{1}>|\tilde
{P}_{i}>$ by $|\tilde {P}_{i}>|\phi_{i}^{1}>$ in the states
$\tilde{
\rho}_{M}^{\cal {E}}$ and $\tilde {\rho}_{E}^{\cal {E}}$.
Noticing that the reverse of the noisy quantum channel leads $\cal {
E}$ to ${\cal {E}}^{\dagger}$, so we should use the
superoperator $\bf {1}\otimes{\cal {E}}^{\dagger}$ as a substitute for
$\cal {E}\otimes \bf {1}$ in the expression of the quantum mutual
information in the backward manner. Based on the fact that the spectra of
general quantum states $\rho$ and $\rho^{\dagger}$ coincide,
we can obviously see that
the quantum mutual information Eq.(5) is invariant under the transformations of
$|\phi_{i}^{1}>|\tilde {P}_{i}>$ into $|\tilde {P}_{i}>|\phi_{i}^{1}>$, and
$\cal {E}\otimes \bf {1}$ into $\bf {1}\otimes {\cal {E}}^{
\dagger}$. From these, we obtain the reverse data processing inequality about
the quantum mutual information, $\tilde {I}_{{\cal {E}}_{2}\circ
{\cal {E}}_{1}}=\tilde {I}_{{\cal {E}}_{1}\circ
{\cal {E}}_{2}}\leq \tilde {I}_{{\cal {E}}_{2}}\leq
S(\tilde {\rho}_{M})$.

The appearance of quantum characteristics in a quantum state is related to
quantum non-separability. In fact, in general, the states of apparatus system
and the states of measured system are not separable. If the general quantum
state
constructed in terms of these states is restricted within the separable case,
i.e., $\rho_{c}=\sum_{i}w_{i}\rho_{i}^{1}\otimes\rho_{i}^{\tilde {P}}$ where
$\rho_{i}^{1}$ is from the Schatten decomposition of the state of the measured
system $\rho_{1}=\sum_{i}w_{i}\rho_{i}^{1}$, we find that the quantum mutual
information Eq.(5) is reduced to $\tilde {I}_{\cal {E}}=S(Tr_{\tilde P}
{\cal E}\otimes {\bf 1}(\rho_{c}))-[S({\cal E}\otimes
{\bf 1}(\rho_{c}))-S(\rho_{c})]=S(Tr_{\tilde P}{\cal E}\otimes
{\bf 1}(\rho_{c}))-\sum_{i} w_{i}S({\cal E}(\rho_i^1))$. The
change of the unmeasured information is present in the square bracket of the
above expression because the separable state $\rho_{c}$ is a quantum mixed
state. The input state can be equivalently described by the "pointer" states of
input exploring the property of quantum measurement. Then, for the case of
separability, our definition of quantum mutual information based on the quantum
measured information deduces to that of Ohya's approach\cite {Ohya} which
is established by means of
the compound state describing the correlation between an input state $\rho_{1}$
and the output state ${\cal {E}}(\rho_{1})$. 
However, we should emphasize that Ohya's approach is not complete
because the quantum coherence of the entanglement states, which plays the
important roles in quantum computation, quantum teleportation and quantum
cryptography, is restricted out in his approach. 

In our scheme, the quantum
mutual information reduces to the classical one when the system is classical.
When the input system is classical, an input state is given by a probability
distribution or a probability measure. For the case of probability distribution, the input state can be expressed by $\rho=\sum_{i}w_{i}\delta_{i}$, where
$\delta$ is the delta measure, i.e., $\delta_{i}(j)=\delta_{i,j}$. As the
special case of the separability, the mutual information for the channel $\cal {
E}$ becomes $I_{{\cal {E}},c}=S({\cal {E}}(\rho))
-\sum_{i} w_{i} S({\cal {E}}(\delta_i))$, which has been taken as the
definition of the mutual information for a classical-quantum-classical channel
\cite {Holevo}.

Since the measure of quantum information is the quantum measured information
in our formalism, we can communicate the same amount of quantum informations
by using the transmissions of the pure state $\rho=
\sum_{i,j} a_{i}a_{j}^{*}|\phi_{i}><\phi_{j}|$ or the mixed state ${\tilde \rho}=\sum_{i} |a_{i}|^{2}|\phi_{i}><\phi_{i}|$ in a Hilbert space. However, in the
process of transmissions, the fidelities for the two type of quantum states
are different. The quantum states $\rho$ and ${\tilde \rho}$ of transmissions
along the noisy quantum channel ${\cal E}$ are evolved into $\rho^{\cal E}=
{\cal E}(\rho)$ and ${\tilde \rho}^{\cal E}={\cal E}({\tilde \rho})$, 
respectively. For the case of transmission of the pure state, its fidelity is
read as $F(\rho,\rho^{\cal E})=Tr(\rho \rho^{\cal E})$. But, the fidelity for 
the mixed quantum states, which is defined in terms of Uhlmann's formula
of transition probability\cite {Jozsa}, is given by $F({\tilde \rho},
{\tilde \rho}^{\cal E})=\{ Tr[(\sqrt {\tilde \rho} {\tilde \rho}^{\cal E}
\sqrt {\tilde \rho} )^{\frac{1}{2}}] \}^{2}$. It is necessary to analyze the
quantitative properties of these fidelities although we do not extensively
discuss this topic here. It is hoped that such analyzing results may be
applicable to developing a quantum analogue of Shannon's channel capacity 
theorem. 

Summarizing, we have introduced the quantum measured information to measure
the quantum information of a quantum state. Using this point of view we have 
consistently decided the quantum mutual information, which measures the amount
of quantum information conveyed in the noisy quantum channel. We has 
proven that such quantum mutual information obeys the data
processing inequality in both forward manner and backward manner.

I thank Z.C. Lu for helpful discussions. The work is supported by 
the NNSF of China (Grant No.19875041), the Special 
NSF of Zhejiang Province (Grant No.RC98022) and Cao Guang-Biao Foundation in 
Zhejiang University.



\begin{references}

\bibitem{Shor}
P. Shor, in Proceedings, 35th Annual Symposium on Foundations of Computer
Science, (IEEE Press, New York, 1994), p.124.

\bibitem{Bennet}
C.H. Bennet, Physical Today, {\it October}, 24 (1995).

R. F. Werner Phys. Rev. A {\bf 40}, 4277 (1989).

\bibitem{DiVincenzo} 
D. P. DiVincenzo, Science {\bf 270}, 255 (1995).

\bibitem{Lloyd}
S. Lloyd, Science {\bf 273}, 1073 (1995).

\bibitem{Bennet1} 
C.H. Bennet, G. Brassard and N.D. Mermin, Phys. Rev. Lett {\bf 68}, 557 (1992).

\bibitem{Bennet2}
C.H. Bennet, G. Brassard, C. Crepeau, R. Jozsa, A. Peres and W.K. Wootters, 
Phys. Rev. Lett {\bf 70}, 1895 (1993).

\bibitem{Schumacher} 
B.W. Schumacher, Phys. Rev. A {\bf 54}, 2614 (1996);
B.W. Schumacher and M.A. Nielson, {\it ibid}. {\bf 54}, 2629 (1996).

\bibitem{Barnum}
H. Barnum, M.A. Nielson and B.W. Schumacher, Phys. Rev. A {\bf 57}, 4153 (1998);H. Barnum, E. Knill and M.A. Nielson, quant-ph/9809010.

\bibitem{Adami}
C. Adami and N.J. Cerf, Phys. Rev. A {\bf 56}, 3470 (1997);
N.J. Cerf and C. Adami, Phys. Rev. Lett. {\bf 79}, 5194 (1997).

\bibitem{Schumacher1}
B.W. Schumacher, Phys. Rev. A {\bf 51}, 2738 (1995);
R. Jozsa and B.W. Schumacher, J. Mod. Opt. {\bf 41}, 2343 (1994).  

\bibitem{Ohya}
M. Ohya, quant-ph/9808051.

\bibitem{Wootters}
W.K. Wootters and W.H. Zurek, Nature (London) {\bf 229}, 802 (1982).

\bibitem{Neumann}
J. von Neumann, Mathematical Foundations of Quantum Mechanics,
( Princeton, New Jersey, 1955 ). 

\bibitem{Wehrl}
A. Wehrl, Rev. Mod. Phys. {\bf 50}, 221 (1978).

\bibitem{Shannon}
S.E. Shannon, Bell system Tech. J. {\bf 27}, 379 (1948).

\bibitem{Holevo}
A.S. Holevo, Prob. Inform. Trans. {\bf 9}, 177 (1973).

\bibitem{Jozsa}
R. Jozsa, J. Mod. Opt. {\bf 41}, 2315 (1994).  

\end{references}
\end{document}